# Lifshitz metal–insulator transition induced by the all-in/all-out magnetic order in the pyrochlore oxide $Cd_2Os_2O_7$

Z. Hiroi,[1,a] J. Yamaura,[2] T. Hirose,[1] I. Nagashima,[1] and Y. Okamoto[1,b]

[1]*Institute for Solid State Physics, University of Tokyo, Kashiwa, Chiba 277–8581, Japan*
[2]*Materials Research Center for Element Strategy, Tokyo Institute of Technology, Yokohama, Kanagawa 226–8503, Japan*

[a]Author to whom correspondence should be addressed. Electronic mail: hiroi@issp.u-tokyo.ac.jp
[b]Present address: Department of Applied Physics, Nagoya University, Chikusa-ku, Nagoya 464–8603, Japan

We investigate the metal–insulator transition (MIT) of the osmium pyrochlore oxide $Cd_2Os_2O_7$ through transport and magnetization measurements. The MIT and a magnetic transition to the all-in/all-out (AIAO) order occur simultaneously at 227 K. We propose a mechanism based on a Lifshitz transition induced by the AIAO magnetic order probably via strong spin–orbit couplings in the specific semimetallic band structure. It is suggested, moreover, that two observed puzzles, a finite conductivity near $T = 0$ and an emergence of weak ferromagnetic moments, are not bulk properties but originate at magnetic domain walls between two kinds of AIAO domains.

## I. INTRODUCTION

The metal–insulator transition (MIT) is one of the most intriguing phenomena observed in various classes of compounds, in which a high-temperature metallic state is transformed into a low-temperature insulating state at a critical temperature ($T_{MI}$).[1] It carries important information about the fundamental property of electrons in crystals, since it is caused by certain electronic instabilities associated with Coulomb interactions among electrons, Fermi-surface (FS) instability, or couplings to other degrees of freedom. A Mott transition is expected for compounds having large electron–electron Coulomb interactions, while a charge- or spin-density wave transition may occur to remove a certain FS instability. Alternatively, antiferromagnetic order alone could produce an insulator by a doubling of a magnetic unit cell in the so-called Slater mechanism.[2] Shedding light on the mechanisms of MITs would help us to understand the underlying properties of "protean" electrons in crystals.

In addition to the academic interest, the MIT is important for technological applications such as sensors and switching devices, because abrupt changes in resistivity and optical absorption at the transition are useful for them.[3,4] For example, the MIT of vanadium dioxide $VO_2$ at $T_{MI} = 340$ K can be utilized as an infrared imaging device operating at room temperature, in which absorbed infrared radiation changes the temperature of a microbolometer made of $VO_2$ with high sensitivity due to a large temperature coefficient of resistance at the transition.[5] Recently, an active control of the $T_{MI}$ has been achieved in thin films or beams by means of electric field or strain.[6,7] In order to improve the performance of any device, a basic understanding of the microscopic mechanism of the MIT is necessary.

The MIT of the osmium pyrochlore oxide $Cd_2Os_2O_7$ has been known for long time since Sleight and coworkers discovered a sudden rise in resistivity below ~225 K in 1974,[8] such as shown in Fig. 1 for our crystal. The compound crystallizes in a cubic pyrochlore structure comprising nearly regular (slightly compressed along the threefold axis) $OsO_6$ octahedra connected by their vertices with interpenetrating Cd–O' framework,[9] as depicted in Fig. 1. The basic electronic properties of the metallic state are determined by three $5d$ electrons in the $t_{2g}$ manifold of the $Os^{5+}$ ions, which form a semimetallic band.[10,11] The most important feature of the band structure near the Fermi level is a small overlap in energy between electron and hole bands at different positions in the $k$ space, which will be addressed later in detail.

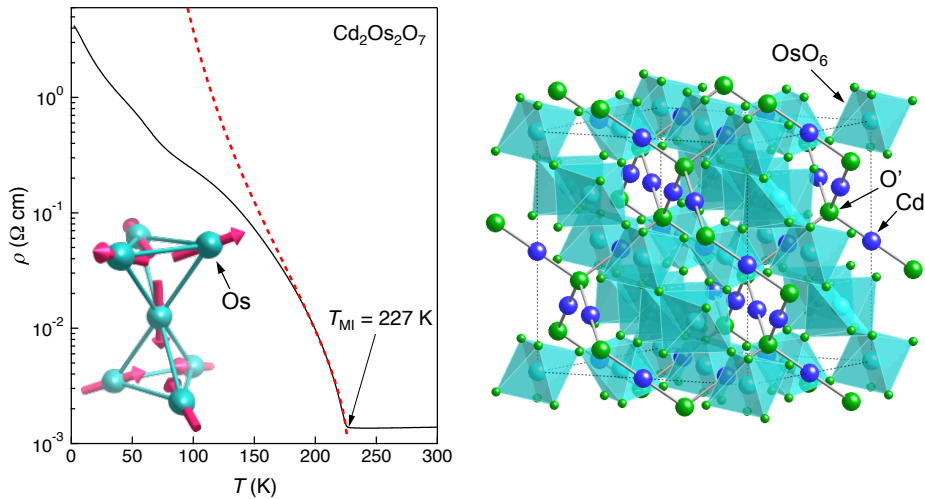

FIG. 1. MIT at $T_{MI}$ = 227 K in resistivity and the AIAO spin arrangement on a pair of Os tetrahedra occurring below $T_{MI}$ (left) and the crystal structure (right) for $Cd_2Os_2O_7$. The red broken curve in the left panel shows a calculated resistivity for a BCS-type gap opening with a zero-temperature gap of 875 K, which has been determined so as to reproduce the data between 190 and 227 K.

Almost quarter-century later after the discovery, Mandrus and coworkers reviewed the MIT of $Cd_2Os_2O_7$ and show that it is a continuous, second-order transition accompanied by a magnetic order.[12] Based on various characterizations, they suggest a Slater-type transition as the mechanism. The following infrared spectroscopy study supports this mechanism, which observes a clear opening of a BCS-type gap with $2\Delta = 5.2k_B T_{MI} \sim 1,200$ K.[13]

More recently, Yamaura and coworkers investigated the spin arrangement below $T_{MI}$ by means of resonant x-ray magnetic scattering on a small single crystal and found a $q = 0$ order that keeps the original cubic primitive cell.[14] They suggest a noncollinear all-in/all-out (AIAO) spin arrangement for the $q = 0$ order, which is a "tetrahedral order"[14, 15] with all the magnetic moments on the four vertices of each Os tetrahedron pointing either inward or outward, as schematically depicted in Fig. 1. This AIAO order has been unequivocally established by the following $^{17}$O NMR experiments by Yamauchi and Takigawa.[16] The AIAO order is unique in the sense that it is regarded as a ferroic order of magnetic octupole moments, which may show peculiar responses to external fields.[17] On the other hand, for related iridium pyrochlore oxides that show similar MITs,[18, 19] it has been also suggested that an AIAO order of Ir moments is realized in their insulating states.[20-23]

The AIAO order is theoretically predicted for $Cd_2Os_2O_7$ by Chern and Batista on a simple double-exchange model without spin–orbit coupling (SOC).[15] However, the importance of SOC has been well recognized in determining the electronic structure[10, 11] and is actually evidenced by means of x-ray magnetic circular dichroism.[24] Shinaoka and coworkers suggest, based on the density-functional theory plus electron correlation, that a large easy-axis anisotropy of 24 meV arising from SOC stabilizes the AIAO order in the presence of a moderate electron correlation of ~1.5 eV.[25] Contrary to this, the $^{17}$O NMR experiments find a power-law temperature dependence close to $T^3$ in the NMR relaxation rate at a wide temperature range of 25–150 K, indicating the lack of a large Ising anisotropy gap in the AIAO state.[16] On the other hand, Bogdanov et al. performed many-body quantum-chemical calculations and find that even a small trigonal distortion of the $OsO_6$ octahedron, as actually observed for $Cd_2Os_2O_7$, causes an easy-axis anisotropy of 6.8 meV, which stabilizes the AIAO order in conjunction with an antiferromagnetic exchange interaction of a similar magnitude of 6.4 meV;[26] this may not be surprising as extended $5d$ orbitals feel very effectively the O $2p$ charge distribution.

Another important finding by Yamaura et al. is the absence of structural changes at the MIT of $Cd_2Os_2O_7$: the cubic space group $Fd$-$3m$ is preserved through the MIT and the magnetic transition with little change in the lattice constant.[14] This imposes a stringent constraint on the mechanism of the MIT since any of mechanisms associated with the Mott, charge/spin-density wave and Slater transitions would require a structural change more or less. What has been suggested for the mechanism based on this in previous study is a kind of Lifshitz transition[27] in the characteristic semimetallic band structure.[14, 25] We would like to address this issue in more detail in the present paper.

In this study, we have carried out transport and magnetization measurements on several high-quality crystals of $Cd_2Os_2O_7$ and try to clarify some important aspects of the transitions. Particularly, a sample dependence is carefully examined. We propose a Lifshitz-type MIT induced by the AIAO magnetic order. It is suggested, moreover, that two experimental puzzles pointed out in previous study, a finite conductivity at $T = 0$ and a weak ferromagnetism, are ascribed to magnetic domain walls (MDWs) in the AIAO order. A marked resemblance in properties between $Cd_2Os_2O_7$ and the related iridium pyrochlore oxides such as $Nd_2Ir_2O_7$, in spite of the substantial difference in the electronic structures, is finally discussed.

## II. EXPERIMENTAL

Single crystals of $Cd_2Os_2O_7$ were grown by the chemical transport method. The process involved two steps. First, a polycrystalline pellet was prepared from a mixture including 5–10% excess CdO and Os in a sealed quartz tube under supply of an appropriate amount of AgO as the oxygen source at 1073 K;[14] addition of too much AgO should be avoided, as highly toxic $OsO_4$ might be produced. In the second step, the pellet was placed at one end of a sealed quartz tube of 30 cm long, which was heated for a week in a furnace having a temperature gradient of 1040–1200 K with the pellet at the high-temperature side. Several crystals of $Cd_2Os_2O_7$ with the truncated octahedral shape of a few mm size at maximum were grown near the middle of the tube, together with crystals of $OsO_2$ and Os at the low-temperature side of the tube. Resistivity and Hall measurements were carried out on picked $Cd_2Os_2O_7$ crystals in a Quantum Design PPMS, and magnetization was measured in a Quantum Design MPMS. A sample dependence in resistivity has been examined on several crystals from different batches prepared under nearly identical but somehow different conditions.

## III. RESULTS AND DISCUSSION

### A. Transition temperatures

First of all, we would like to focus on the transition temperatures. As shown in Fig. 2, which is the blowup of Fig. 1, there is a clear kink in resistivity for crystal A at 227.2 K, where the temperature derivative bends over sharply. Thus, we can determine $T_{MI} = 227.2$ K unequivocally, which is close to 225 K by Sleight et al.[8] and 226 K by Mandrus et al.[12, 13] We examined several crystals from different preparation batches by resistivity and found a small sample dependence with $T_{MI}$ = 227–229 K, indicating little deviation in stoichiometry or an insensitivity of the $T_{MI}$ against certain perturbations.

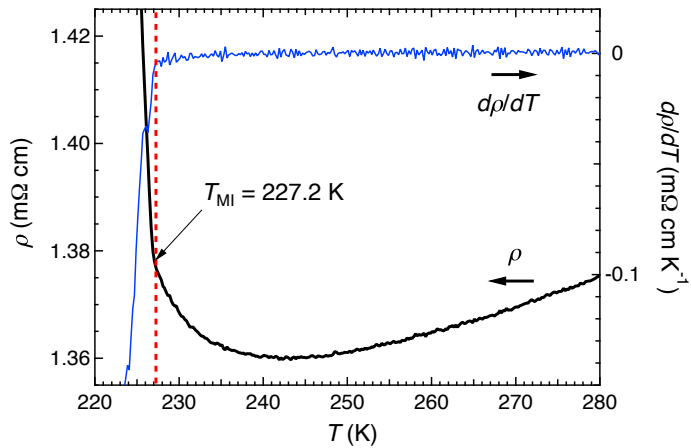

FIG. 2. Blowup of the resistivity of crystal A shown in Fig. 1 and its temperature derivative. Clear kinks are observed at $T_{MI} = 227.2$ K in both curves.

The magnetic susceptibility of a polycrystalline sample shown in Fig. 3 exhibits a clear cusp at 227 K in the zero-field cooling (ZFC) curve, below which the field cooling (FC) curve separates from the ZFC curve, resulting in a thermal hysteresis. The presence of the sharp cusp means that a distribution in the transition temperature is negligible among a number of crystallites contained in the sample. Thus, a magnetic order sets in at $T_{mag} = 227$ K. These resistivity and magnetic susceptibility data reveal that the MIT and the magnetic transition are really concomitant with each other. In addition, the $^{17}$O NMR experiments show an almost equal magnetic transition temperature of 227.19(4) K, below which the internal magnetic field grows with a marked second-order character.[16]

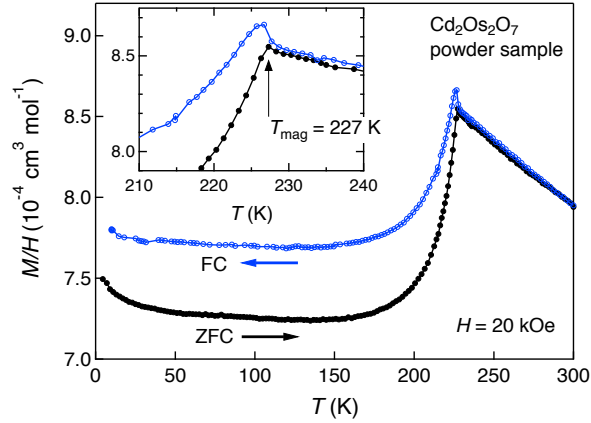

FIG. 3. Temperature dependences of magnetization $M$ divided by magnetic field $H$ measured on a polycrystalline sample. The measurements are carried out first upon heating at $H$ = 20 kOe after cooling in zero field (ZFC) and then upon cooling in the same field (FC). A cusp indicative of a magnetic long-range order is observed at 227 K.

## B. Transport properties

One notable feature in the resistivity of Fig. 2 is that it already begins to increase below a shallow minimum at around 242 K, 15 K higher than the $T_{MI}$. There are two possibilities to explain this: one is due to an enhanced magnetic scattering of carriers in approaching the magnetic order, and the other is associated with a reduction in carrier density. The former case is sometimes found for metallic frustrated spin systems. For example, in a related pyrochlore oxide $Pr_2Ir_2O_7$,[28] a resistivity minimum is observed at ~50 K and is ascribed to a magnetic scattering due to local spin-ice correlations under geometrical frustration.[29] In $Cd_2Os_2O_7$, in contrast, frustration effects may be unimportant, judging from the high ordering temperature. In addition, we observed almost no magnetoresistance up to 9 T in this temperature range. Therefore, the origin of the resistivity minimum must be related to a reduction in carrier density, as will be discussed later.

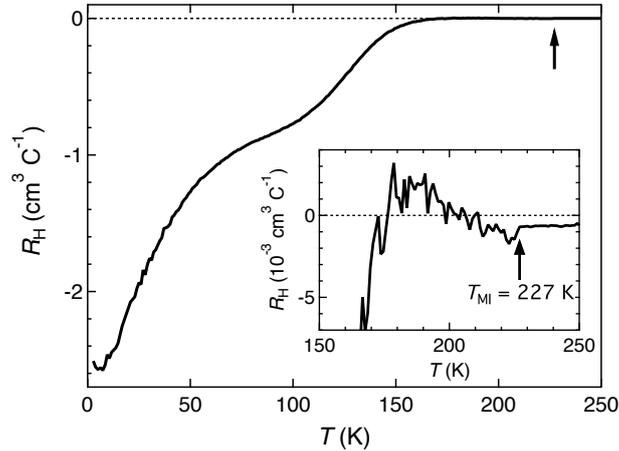

FIG. 4. Hall coefficient $R_H$ measured on a crystal of $Cd_2Os_2O_7$, which was obtained by measuring Hall voltages in $H$ = ±90 kOe at each temperature. The inset shows a blowup around $T_{MI}$ = 227 K.

In order to get information on the carrier density, Hall measurements were performed on a single crystal prepared in the same condition as crystal A. As shown in Fig. 4, the Hall coefficient $R_H$ shows an anomaly at 227 K at the MIT. Above the $T_{MI}$, it takes a small, $T$-independent negative value of about $-7 \times 10^{-4}$ cm$^3$ C$^{-1}$. Unfortunately, we cannot deduce the carrier density from this value since the compound is ideally a compensated metal with the same density of electrons ($n_e$) and holes ($n_h$). For a compensated metal, the conductivity and the Hall coefficient are give by

$$\sigma = q(n_e\mu_e + n_h\mu_h), \quad (1)$$
$$R_H = \frac{n_h\mu_h^2 - n_e\mu_e^2}{q(n_e\mu_e + n_h\mu_h)^2}, \quad (2)$$

where $q$ is the elementary charge, and $\mu_e$ and $\mu_h$ are mobilities of electrons and holes, respectively. The observed small negative value means either that $\mu_e$ is slightly larger than $\mu_h$, or that $n_e$ is slightly larger than $n_h$ owing to the presence of additional electrons possibly from impurity states: $n_e = n_h + n_{imp}$. Taking values of $R_H = -7.0\times10^{-4}$ cm$^3$ C$^{-1}$ and $\rho = 1.36\times10^{-3}$ $\Omega$ cm at 230 K, the difference in the mobility for the case of $n_e = n_h$ is deduced to be 0.52 cm$^2$ V$^{-1}$ s$^{-1}$. Provided that $n_e = n_h = n_{band} = 1.27\times10^{21}$ cm$^{-3}$,[11] which corresponds to 0.08 per Os, the $\mu_e$ and $\mu_h$ are 2.08 and 1.56 cm$^2$ V$^{-1}$ s$^{-1}$, respectively. On the other hand, assuming $n_{imp} = 6.6\times10^{18}$ cm$^{-3}$ from the estimation described below would give $n_h = 1.24\times10^{20}$ cm$^{-3}$ for $\mu_e = \mu_h$, about one order smaller than $n_{band}$: the assumption of $\mu_e = \mu_h$ might be inappropriate. Thus, a small unbalance in the magnitudes of mobility between electrons and holes may govern the $R_H$ above $T_{MI}$.

Upon cooling below $T_{MI}$, the $R_H$ becomes positive and then negative, followed by a steep decrease below 170 K. Interestingly, it tends to saturate at 50–100 K and then reaches another saturation at a larger negative value toward $T = 0$; previously reported $R_H$ on a polycrystalline sample has a similar but slightly different temperature dependence and a smaller magnitude by about a factor of two.[12] These "plateaus" indicate that some carriers of different origins have shown up after the major electrons and holes are diminished with the opening of a gap. The $R_H$ values at the two "plateaus" could be translated into excess electron carrier densities, if one assumes a single carrier model: $R_H = -0.95$ cm$^3$ C$^{-1}$ at 75 K and $-2.5$ cm$^3$ C$^{-1}$ at 4 K correspond to $n_{imp} = 6.6\times10^{18}$ cm$^{-3}$ and $2.5\times10^{18}$ cm$^{-3}$, respectively. In addition, from $\rho = 0.38$ $\Omega$ cm at 75 K for crystal A in Fig. 1, the mobility at 75 K is obtained as 2.5 cm$^2$ V$^{-1}$ s$^{-1}$, which is close to those estimated at 230 K. More implications from these transport data will be addressed later.

## C. Lifshitz MIT

The mechanism of the MIT of $Cd_2Os_2O_7$ is no doubt related to its semimetallic band structure with an indirect band gap. According to the band structure calculations by two groups,[10,11] there is a $t_{2g}$ manifold consisting of 12 bands (per primitive cell including four Os atoms) near the Fermi level, which is exactly half filled. However, since the total bandwidth of ~3 eV is larger than the effective Hubbard $U$ of 1–2 eV, $Cd_2Os_2O_7$ should not be classified as a strongly correlated material,[10] and the SOC physics may prevail for this 5$d$ electron system. To be focused are sixth, seventh and eighth bands that cross the Fermi level: the sixth band generates hole-like FSs at the zone boundary surrounding the W points, while the other two bands form two electron-like FSs surrounding the $\Gamma$ point and along the $\Gamma$–X line. The band structure of $Cd_2Os_2O_7$ is schematically depicted in Fig. 5, where the minor electron band is omitted for clarity. Singh and coworkers have pointed out that the band structure can be, at least conceptually, made insulating by removing the overlap in energy between the electron and hole bands by some way.[10]

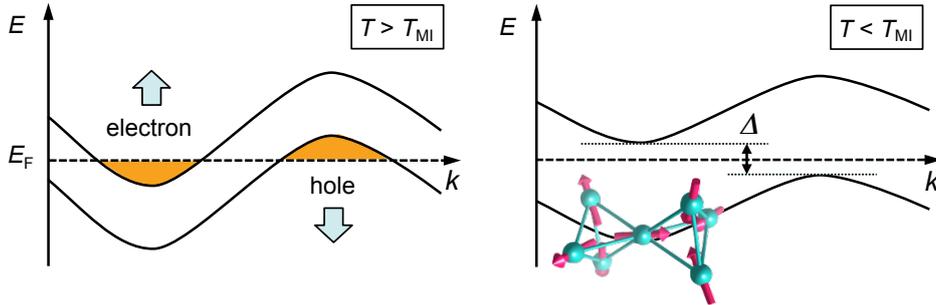

FIG. 5. Schematic representations of simplified band structures expected for a paramagnetic, semimetallic state above $T_{MI}$ (left) and for an antiferromagnetic, insulating state below $T_{MI}$ (right). The overlap of approximately 120 meV in energy between the electron and hole bands in the former is going to be removed by upward and downward shifts of the electron and hole bands, respectively, or by a narrowing of each band, which may be caused by the AIAO magnetic order via strong SOCs. The mechanism of this MIT is considered as a magnetic order-induced Lifshitz transition of the pocket-vanishing type.

The two experimental facts, the absence of structural changes and the emergence of the $q = 0$ magnetic order without Brillouin-zone folding, indicate that such an FS-vanishing transition actually takes place in $Cd_2Os_2O_7$. It is considered as a sort of Lifshitz transition that is an electronic transition due to the variation of the FS topology.[27] Pressure-induced transitions observed in some compounds have been ascribed to either pocket-vanishing or neck-collapsing type Lifshitz transitions.[30] Moreover, a pocket-vanishing type Lifshitz transition has been observed in $Na_xCoO_2$ as a function of carrier density.[31] Note that all the known Lifshitz transitions are metal–metal transitions, while the present compound possibly gives a first example of a Lifshitz-type metal–insulator transition, where all the FSs are gone.

One notable feature on the band structure of $Cd_2Os_2O_7$ that makes the Lifshitz MIT possible is the small overlap in energy between the electron and hole bands: according to the band structure calculations, the MIT can occur when the bottom or top of each band moves by ~60 meV to get rid of the overlap between them.[10] It is worthwhile referring to another pyrochlore compound $Hg_2Os_2O_7$ which has essentially the same electron configuration as $Cd_2Os_2O_7$ but seems to remain a metal down to 4 K.[32] According to band structure calculations by Harima, $Hg_2Os_2O_7$ has basically the same band structure as $Cd_2Os_2O_7$, but the overlap in energy between the hole and electron bands is approximately 200 meV, larger than 120 meV for $Cd_2Os_2O_7$.[33] Thus, one would expect that a larger energy gain is required for a Lifshitz MIT to occur in $Hg_2Os_2O_7$, which may be the reason for the lack of MIT.

A driving force for the Lifshitz MIT of $Cd_2Os_2O_7$ must be the AIAO magnetic order, because the MIT and the magnetic order occur simultaneously. It is expected in Fig. 5 that, when the electron band shifts upward and the hole band shifts downward, or when both bands become narrower in the semimetallic band structure, their overlap in energy eventually becomes zero just at $T_{MI}$; since it is a continuous transition, there is no jump but a kink in resistivity at $T_{MI}$, as observed in Fig. 2. At the same time, a magnetic order arising from the sinking $t_{2g}$ band sets in at $T_{MI}$. In order to make this coincidence possible, the band movement or narrowing should have already started at higher temperatures above $T_{MI}$ and is enhanced with increasing magnetic fluctuations toward the long-range order. This is partially evidenced by the observed upturn in resistivity below 242 K in Fig. 2: the carrier density has been already decreasing with the band movements. In this sense, the MIT of $Cd_2Os_2O_7$ is assumed to be a magnetic Lifshitz MIT or a Lifshitz MIT induced by a magnetic order. It is crucial to figure out why such band shifts or narrowings are induced by the magnetic order. Intuitively plausible is that a strong SOC adjusts the bands to open a gap, when the magnetic order is approached. A microscopic theory would be required for further discussion.

### D. One mystery: finite conductivity at $T = 0$

Next, we would like to address two mysteries raised in previous study. One is the unusual temperature dependence in resistivity below $T_{MI}$, which is followed by a saturation near $T = 0$ even in the "insulating" state. The resistivity of crystal A shown in Fig. 1 sharply rises below $T_{MI}$ as expected for an opening of a BCS-type gap of 875 K (shown by a broken line), but the rise is significantly suppressed at low temperatures, followed by a finite value of 4 Ω cm at 4 K; similar behavior has been reported on single crystals and polycrystalline samples by three groups.[8, 12, 14] In contrast, optical measurements evidence a conventional opening of a gap of a similar magnitude.[13]

In order to get insight on this issue, we have carefully examined a sample dependence in resistivity on five crystals picked up from same or different preparation batches (Fig. 6). It is noticed that the initial rises of their resistivities are more or less similar to each other. The corresponding activation energy $\Delta(T)$ is calculated from the resistivity data using $\Delta(T) = T\ln(\rho/\rho_0)$.[12] The $\Delta(T)$s of crystals A and D are approximately reproduced by BCS gap functions[34] with zero-temperature gaps of $\Delta(0) = 670$ and 780 K, respectively, as shown in Fig. 6(b). Whether the zero-temperature gap is actually variable or not is open to question, because some influence to conductivity from other contributions mentioned below is already expected even near $T_{MI}$, which may cause an underestimation for the magnitude of the gap.

In sharp contrast, a large scatter in the $T$ dependence appears below ~150 K: compared with crystal A, crystals B and C/C' show larger and smaller increases, respectively, and crystal D exhibits even metallic behavior at an intermediate $T$ range of 70–130 K! This marked sample dependence strongly suggests that the resistivity in this $T$ range is not a bulk characteristic but is governed by impurity conduction; the amount of impurities in crystals could vary depending on preparation conditions. Note that the resistivities of the two crystals C and C' from one batch resemble each other. One evidence to support a contribution from impurity conduction is found in the resistivity of crystal D, which shows a steep increase below ~50 K that follows an activation type with a small activation energy of 100 K in the $T$ range of 20–40 K. Thus, certain impurity states must exist inside the gap. The impurity conduction manifests itself at low temperatures when thermally excited carries across the main BCS-type gap are diminished with increasing the gap upon cooling, and finally should disappear at $T = 0$. Besides this, the $R_H$ of Fig. 4 tends to saturate to a large negative value in this $T$ range, indicating that what dominates the transport there are extra electrons with an approximate density of $n_{imp} = 6.6\times10^{18}$ cm$^{-3}$ from donor levels 100 K below the conduction band. Possibly, the crystals contain small amounts of oxygen vacancy, as they were prepared in slightly reducing atmosphere, and the amounts could vary from crystal to crystal depending on preparation conditions. The amount of oxygen vacancy to account for the $n_{imp}$ value is really tiny, 0.0004 per formula unit, which is too small to detect by any chemical analysis or structural refinement. A more detailed analysis on the impurity conduction will be given elsewhere.[35]

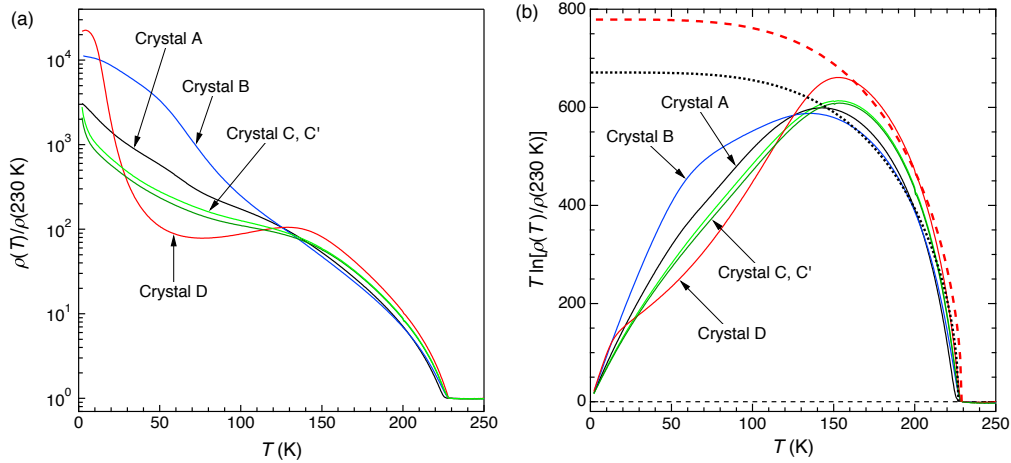

FIG. 6. (a) Sample dependence in normalized resistivity. The reference of the normalization is set to values at 230 K slightly above $T_{MI}$, which are $\rho_0 = 0.5$–1.3 m$\Omega$ cm for these crystals; the raw data of crystal A is shown in Fig. 1. Crystals A, B, C and D are from different batches, and crystals C and C' are from a same batch. (b) Temperature dependences of the activation energy $\Delta(T)$ calculated from the resistivity data using $\Delta(T) = T\ln(\rho/\rho_0)$. The black dotted and red broken curves show fits to the BCS gap functions with $\Delta(0) = 670$ and 780 K for crystals A and D, respectively.

At the lowest temperature, the resistivity curves are going to saturate completely to large values in crystals A, B and D or are terminated with small upturns in crystals C and C'; the small upturn is probably due to the weak localization, as evidenced by its characteristic magnetic field dependence.[35] Particularly, the resistivity of crystal D shows a clear saturation below ~10 K after the steep rise below 50 K due to the extinction of impurity-induced electron carriers; note that the crystal is not in the regime of the degenerate semiconductor with a large density of impurity states. Thus, this finite conductivity at $T = 0$ cannot be interpreted as a bulk property: the bulk must be highly insulating at the lowest temperature as evidenced typically by the optical measurements.[13] Therefore, it is reasonable to presume that local conducting paths exist somewhere in a crystal. The saturation of the $R_H$ at the lowest temperature in Fig. 4 may also be due to the conducting paths.

### E. Another mystery: weak ferromagnetic moments

Another mystery is the appearance of weak ferromagnetic (WF) moments below $T_{MI}$, as observed in the thermal hysteresis between the ZFC and FC curves in Fig. 3: randomly oriented ferromagnetic moments in the ZFC state are forced to align partially or completely along the applied magnetic field in the FC process, which results in a thermal hysteresis in $M/H$; this thermal hysteresis is not related to spin glassiness because the well-defined long-range order sets in at the same time. The magnitude of the WF moment is very small compared with the net Os moment of ~1 $\mu_B$ from the NMR study:[16] it would correspond to approximately $2\times10^{-4}$ $\mu_B$ per Os for a large crystal, as shown in Fig. 7. Thus, they must be parasitic due to a tiny spin canting or not from a bulk property. A spin canting is not likely, because the AIAO order should not occur concomitant with spin canting: according to the representation analysis for the $q = 0$ order, the basis function $\Psi_1$ corresponding to the AIAO spin arrangement on one Os tetrahedron cannot be mixed with ferromagnetic basis functions which belong to other irreducible representations, provided that the transition is of the second order.[14] In addition, an Ising anisotropy along the local [111] direction would make a spin canting unfavorable.[25, 26] Therefore, the WF moments should find their origin somewhere else.

Figure 7 shows magnetizations of five samples as a function of magnetic field, which were measured at 190 K below $T_{MI}$ with each measurement after field cooling from above $T_{MI}$. It was not easy to obtain a reliable *M–H* curve in a conventional way with varying magnetic field at each temperature because of the strong anisotropy. This is quite unusual for normal ferromagnets and suggests that the directions of the WF moments are fixed by the isotropic AIAO order which is insensitive to a magnetic field. Noted in Fig. 7 is a significant size dependence of the magnitude of WF moments: they are $2\times10^{-4}$ $\mu_B$ per Os for one large crystal of a few mm size, about half after crushing into small pieces or for ~20 small crystals, and about 1/10 for a polycrystalline sample. It seems that the WF moments are reduced with decreasing the size of crystallites. Thus, they are not uniformly distributed in a crystal. On the other hand, the magnitude becomes more than twice as large for an Ir-doped crystal (we intentionally substituted Ir for Os, but do not know how much has been actually introduced). We have also performed a systematic Re substitution for Os in a series of polycrystalline samples and observed a larger enhancement in the WF moment, i.e. a five times enlargement for a 3% Re substitution. These size and doping dependences strongly suggest that the WF moments are not located at the crystal surface (the powder sample with large surface area has the smallest value) but at magnetic domain walls (MDWs); the density of MDWs may be sensitive to the size of crystals and can be increased by introducing impurities, as they serve to pin down MDWs.

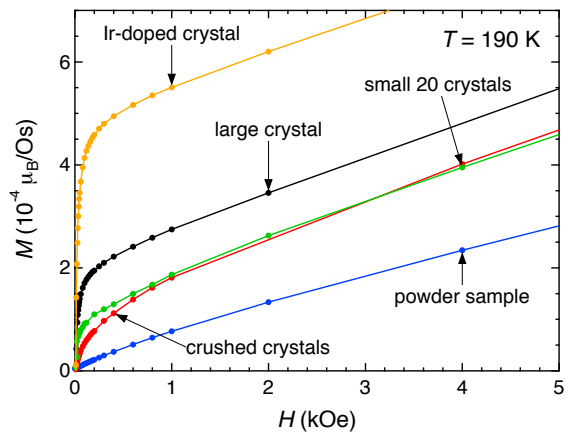

FIG. 7. *M–H* curves of various samples at 190 K, which were measured after field cooling from above $T_{MI}$ in each field. The WF moment of $\sim2\times10^{-4}$ $\mu_B$ per Os for one large crystal of a few millimeter size is reduced nearly half after crushing into small pieces or for 20 smaller crystals, and more reduced for a polycrystalline sample. In contrast, it is enhanced more than double for an Ir-for-Os substituted crystal.

**F. Magnetic domain walls**

As noted by Arima,[17] since the AIAO order is a unique ferroic order of magnetic octupole moments on Os tetrahedra, it shows bistability, just as the conventional ferromagnetic order, in spite of the absence of a net moment. Two kinds of magnetic domains related by time-reversal symmetry can coexist under certain conditions, one of which has the all-in/all-out spin arrangement and the other has the all-out/all-in spin arrangement, as schematically depicted in Fig. 8. Thus, MDWs between them should occur when time-reversal symmetry is broken below $T_{MI}$. However, they ought to disappear at $T = 0$, because the generation of MDWs costs some energy. In reality, as usually the case for ferromagnets, they must persist as being trapped by crystalline defects or by other extrinsic effects. Very recently, a synchrotron x-ray imaging technique using right- and left-handed circular polarization beams has made it possible to observe such MDWs in $Cd_2Os_2O_7$:[36] they appear with approximately 50 μm spacing on the surface of a crystal at 100 K after cooling in a magnetic field. Moreover, it is demonstrated that external magnetic fields can control the distribution of the MDWs to some extent.

Let us consider a possibility that certain MDWs can really carry uncompensated magnetic moments in the classical local spin picture. Among possible MDWs, here we consider a (100) MDW. When two domains meet with each other at a (100) plane, one expects that the interface layer contains Os tetrahedra with a two-in/two-out spin arrangement that carry a net moment fixed perpendicular to the (100) plane, as sketched in Fig. 8. Moreover, all the uncompensated magnetic moments at the MDWs point right or left in the same direction, if every domain contains equal numbers of two kinds of tetrahedra with all-in and all-out spin arrangements so as not to break the magnetic unit cell of the AIAO order. To obtain the experimental magnitude of the WF moment of $2\times10^{-4}$ $\mu_B$ per Os by taking only this kind of MDWs, it is enough to assume that MDWs, each carrying a net moment of 1 $\mu_B$ per Os, appear with a few $\mu$m separation. This local spin picture may be too simple to expound the actual situation, because the compound exists near the metal–insulator boundary: an extended electron picture might be more appropriate to assume, which takes into account possible modulations in the magnitude of magnetic moments near MDWs.[37]

The observed finite conductivity near $T = 0$ may be also ascribed to MDWs. Recently, we have observed that resistivity only below ~20 K is markedly influenced by cooling processes in magnetic fields, which will be reported elsewhere.[35] It is plausible that a contribution of conducting MDWs emerges only at low temperatures when the bulk is rendered highly insulating with increasing the BCS-type gap and also when impurity-induced carriers are diminished. Since the density of metallic MDWs must be small at low temperatures, their contribution to conductivity is rather limited, so that the apparent resistivity stays large but its temperature dependence is that of metals. Note that, if the AIAO magnetic order really causes such a band separation as shown in Fig. 5, the interface layer having different spin arrangements such as two-in/two-out could remain metallic with a band overlap surviving. Therefore, we speculate that the MDWs generate ferromagnetic and metallic interface layers, which can give us persuasive explanations to various experimental results but needs to be examined by further experiments and theoretical considerations.

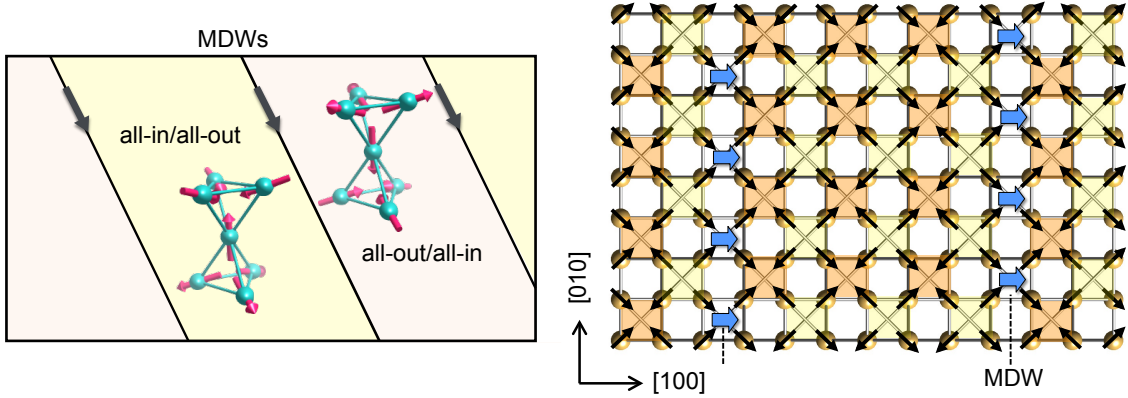

FIG. 8. Schematic drawings of magnetic domain walls (MDWs) generated between two kinds of AIAO domains with all-in/all-out and all-out/all-in spin arrangements: a macroscopic crystal with MDWs (left) and an atomic structure around one type of MDWs with the (100) interface (right), where two kinds of Os tetrahedra viewed along the [001] direction are shaded with orange (all-in) and yellow (all-out) colors. A typical separation of MDWs may be a few $\mu$m, but the separation must depend on various factors such as temperature, magnetic field, field-cooling process, density of defects and crystal size. The two (100) MDWs in the right picture are set at a short separation so as to show the relation between them, which include tetrahedra with two-in/two-out spin arrangements having uncompensated magnetic moments (large arrows) perpendicular to the interface. MDWs may also remain conducting even at $T = 0$.

## G. Relevance to the iridium pyrochlore oxides

Finally, we would like to point out striking similarities in properties between the two 5$d$ pyrochlore oxides, $Cd_2Os_2O_7$ and $R_2Ir_2O_7$ (R represents one of rare earth elements other than La, Pr and Ce),[18, 19] in spite of apparent difference in the basic electronic structures between $Os^{5+}$ (5$d^3$) and $Ir^{4+}$ (5$d^5$). For example, $Nd_2Ir_2O_7$ undergoes an MIT at 33 K, which seems to be accompanied by an AIAO magnetic order with a small ferromagnetic moment of ~5×10$^{-3}$ $\mu_B$ per Ir.[19] Moreover, in $Nd_2Ir_2O_7$, the resistivity tends to saturate below ~30 K, and anomalous magnetoresistances are observed at lower temperatures,[22, 38, 39] which have been attributed to conducting domain walls.[22]

The weak ferromagnetic moments and the remaining conductivity of $Nd_2Ir_2O_7$ have been discussed in terms of the MDWs of the AIAO order,[37] just as mentioned in the present manuscript for $Cd_2Os_2O_7$. Nevertheless, a very different physics behind these observations is under consideration for $R_2Ir_2O_7$: $R_2Ir_2O_7$ is considered as a candidate of an unconventional topological quantum state called the Weyl semimetal.[40-42] The Weyl semimetal is theoretically predicted to have a peculiar metallic state with a Fermi arc at the surface. In addition, Yamaji and Imada note that ferromagnetic and metallic MDWs can occur in the AIAO magnetic order and survive even at low temperatures after the metallic surface state has been diminished by pair annihilation of the Weyl electrons, which may solve the experimental puzzles for $R_2Ir_2O_7$.[37]

In striking contrast, the band structure of $Cd_2Os_2O_7$ probably lacks such features as the Weyl semimetal has, and thus the same interpretation would be unrealistic; $Cd_2Os_2O_7$ seems to be classified into another class of quantum phase called "Axion insulator",[42] though implications from this are not clear. Nevertheless, many similar features observed for the two $5d$ pyrochlore oxides make us believe a possibility that there is a common origin free from specific band structures.

## IV. CONCLUSION

In summary, we have studied the MIT of $Cd_2Os_2O_7$ through transport and magnetization measurements on single crystals. It is shown that the MIT and the AIAO magnetic order take place at exactly the same temperature, suggesting that the MIT is induced by the magnetic order via a Lifshitz transition in the semimetallic band structure with a small overlap in energy between the electron and hole bands. It is also suggested that the magnetic domain walls in the AIAO order carry ferromagnetic moments and remain conducting even at $T = 0$.

## ACKNOWLEDGMENTS

We are grateful to H. Harima for detailed information about the band structures of $Cd_2Os_2O_7$ and $Hg_2Os_2O_7$, and also for helpful discussion. We thank M. Takigawa, Y. Motome, K. Matsuhira, and T. Arima for valuable comments on the manuscript.## REFERENCES

[1] M. Imada, A. Fujimori, and Y. Tokura, Rev. Mod. Phys. **70**, 1039 (1998).
[2] J. C. Slater, Phys. Rev. **82**, 538 (1951).
[3] H. Jerominek, F. Picard, and D. Vincent, Opt. Eng. **32**, 2092 (1993).
[4] A. Cavalleri, C. Toth, C. W. Siders, and J. A. Squier, Phys. Rev. Lett. **87**, 237401 (2001).
[5] M. Gurvitch, S. Luryi, A. Polyakov, and A. Shabalov, J. Appl. Phys. **106**, 101504 (2009).
[6] M. Nakano, K. Shibuya, D. Okuyama, T. Hatano, S. Ono, M. Kawasaki, Y. Iwasa, and Y. Tokura, Nature **487**, 459 (2012).
[7] J. Cao, E. Ertekin, V. Srinivasan, W. Fan, S. Huang, H. Zheng, J. W. L. Yim, D. R. Khanal, D. F. Ogletree, J. C. Grossman, and J. Wu, Nat Nano **4**, 732 (2009).
[8] A. W. Sleight, J. L. Gilson, J. F. Weiher, and W. Bindloss, Solid State Commun. **14**, 357 (1974).
[9] M. A. Subramanian, G. Aravamudan, and G. V. S. Rao, Prog. Solid State Chem. **15**, 55 (1983).
[10] D. J. Singh, P. Blaha, K. Schwarz, and J. O. Sofo, Phys. Rev. B **65**, 155109 (2002).
[11] H. Harima, J. Phys. Chem. Solids **63**, 1035 (2002).
[12] D. Mandrus, J. R. Thompson, R. Gaal, L. Forro, J. C. Bryan, B. C. Chakoumakos, L. M. Woods, B. C. Sales, R. S. Fishman, and V. Keppens, Phys. Rev. B **63**, 195104 (2001).
[13] W. Padilla, D. Mandrus, and D. Basov, Phys. Rev. B **66**, 035120 (2002).
[14] J. Yamaura, K. Ohgushi, H. Ohsumi, T. Hasegawa, I. Yamauchi, K. Sugimoto, S. Takeshita, A. Tokuda, M. Takata, M. Udagawa, M. Takigawa, H. Harima, T. Arima, and Z. Hiroi, Phys. Rev. Lett. **108**, 247205 (2012).
[15] G.-W. Chern, and C. Batista, Phys. Rev. Lett. **107**, 186403 (2011).
[16] I. Yamauchi, and M. Takigawa, in preparation.
[17] T. Arima, J. Phys. Soc. Jpn. **82**, 013705 (2013).
[18] K. Matsuhira, M. Wakeshima, R. Nakanishi, T. Yamada, A. Nakamura, W. Kawano, S. Takagi, and Y. Hinatsu, J. Phys. Soc. Jpn. **76**, 043706 (2007).
[19] K. Matsuhira, M. Wakeshima, Y. Hinatsu, and S. Takagi, J. Phys. Soc. Jpn. **80**, 094701 (2011).
[20] K. Tomiyasu, K. Matsuhira, K. Iwasa, M. Watahiki, S. Takagi, M. Wakeshima, Y. Hinatsu, M. Yokoyama, K. Ohoyama, and K. Yamada, J. Phys. Soc. Jpn. **81**, 034709 (2012).
[21] H. Sagayama, D. Uematsu, T. Arima, K. Sugimoto, J. J. Ishikawa, E. O'Farrell, and S. Nakatsuji, Phys. Rev. B **87**, 100403 (2013).
[22] K. Ueda, J. Fujioka, Y. Takahashi, T. Suzuki, S. Ishiwata, Y. Taguchi, M. Kawasaki, and Y. Tokura, Phys. Rev. B **89**, 075127 (2014).
[23] S. M. Disseler, Phys. Rev. B **89**, 140413 (2014).
[24] Y. H. Matsuda, J. L. Her, S. Michimura, T. Inami, M. Suzuki, N. Kawamura, M. Mizumaki, K. Kindo, J.10

Yamauara, and Z. Hiroi, Phys. Rev. B **84**, 174431 (2011).
[25]H. Shinaoka, T. Miyake, and S. Ishibashi, Phys. Rev. Lett. **108**, 247204 (2012).
[26]N. Bogdanov, R. Maurice, I. Rousochatzakis, J. van den Brink, and L. Hozoi, Phys. Rev. Lett. **110**, 127206 (2013).
[27]I. M. Lifshitz, Sov. Phys. JETP **11**, 1130 (1960).
[28]S. Nakatsuji, Y. Machida, Y. Maeno, T. Tayama, T. Sakakibara, J. v. Duijn, L. Balicas, J. N. Millican, R. T. Macaluso, and J. Y. Chan, Phy. Rev. Lett. **96**, 087204 (2006).
[29]M. Udagawa, H. Ishizuka, and Y. Motome, Phys. Rev. Lett. **108**, 066406 (2012).
[30]B. Godwal, A. Jayaraman, S. Meenakshi, R. Rao, S. Sikka, and V. Vijayakumar, Phys. Rev. B **57**, 773 (1998).
[31]Y. Okamoto, A. Nishio, and Z. Hiroi, Phys. Rev. B **81**, 121102 (2010).
[32]J. Reading, S. Gordeev, and M. T. Weller, J. Mater. Chem. **12**, 646 (2002).
[33]H. Harima, private communication.
[34]T. Sheahen, Phys. Rev. **149**, 368 (1966).
[35]T. Hirose, J. Yamaura, and Z. Hiroi, in preparation.
[36]S. Tardif, S. Takeshita, H. Osumi, J. Yamaura, D. Okuyama, Z. Hiroi, M. Takata, and T. Arima, arXiv:1407.5401 (2015).
[37]Y. Yamaji, and M. Imada, Phys. Rev. X **4**, 021035 (2014).
[38]K. Matsuhira, M. Tokunaga, M. Wakeshima, Y. Hinatsu, and S. Takagi, J. Phys. Soc. Jpn. **82**, 023706 (2013).
[39]S. Disseler, S. Giblin, C. Dhital, K. Lukas, S. Wilson, and M. Graf, Phys. Rev. B **87**, 060403 (2013).
[40]X. Wan, A. Turner, A. Vishwanath, and S. Savrasov, Phys. Rev. B **83**, 205101 (2011).
[41]W. Witczak-Krempa, and Y. Kim, Phys. Rev. B **85**, 045124 (2012).
[42]W. Witczak-Krempa, G. Chen, Y. B. Kim, and L. Balents, Ann. Rev. Cond. Mat. Phys. **5**, 57 (2014).